\documentclass[journal]{IEEEtran}

\usepackage{comment}
\renewcommand{\baselinestretch}{0.92}
\usepackage{changepage}
\usepackage{amsmath}
\usepackage{amssymb}
\usepackage[dvips]{graphicx}
\usepackage{setspace}
\usepackage{epsfig}
\usepackage{color}
\usepackage{morefloats}
\usepackage{enumerate}
\usepackage{bbm}
\usepackage{cases}
\usepackage{enumitem}

\usepackage{graphicx}
\usepackage[caption=false]{subfig}
\allowdisplaybreaks
\newcommand{\mtA}{{\mathcal{A}}}

\newcommand{\PL}{{\rm{PL}}}
\newcommand{\EE}{{\rm{EE}}}

\newcommand{\regret}{\Delta_{\rm{reg}}}
\newcommand{\gNB}{\rm{gNB}}

\usepackage[a4paper,
            bindingoffset=0.2in,
            left=0.6in,
            right=0.6in,
            top=0.7in,
            bottom=0.9in,
            footskip=.25in]{geometry}

\begin{document}
\renewcommand{\baselinestretch}{0.85}
% \pagestyle{empty}
% Before the part you want to change

\title{Energy-Efficient Sleep Mode Optimization of 5G mmWave Networks Using Deep Contextual MAB}
%Optimal Channel Switching and Randomization over Flat-Fading Channels for \textcolor{blue}{Outage} Capacity Maximization
\author{Ismail Guvenc\vspace{-0.5cm}}
\author{Saad Masrur$^1$, \.{I}smail G\"{u}ven\c{c}$^1$, David L\'opez-P\'erez$^2$\\
$^1$Department of Electrical and Computer Engineering, North Carolina State University, Raleigh, NC\\
$^2$Universitat Polit\`ecnica de Val\`encia, Valencia, Spain\\
{\tt \{smasrur,iguvenc\}@ncsu.edu}

\thanks{This research is supported in part by the NSF project CNS 1910153, the Generalitat Valenciana through the CIDEGENT PlaGenT, Grant CIDEXG/2022/17, Project iTENTE, and by the action CNS2023-144333, financed by MCIN/AEI/10.13039/501100011033 and the European Union “NextGenerationEU”/PRTR.”}}

% \author{{Saad Masrur\thanks{S. Masrur and S. Gezici are with the Dept. of Electrical and Electronics Eng., Bilkent Univ., Bilkent, Ankara, Turkey, Tel: +90 (312) 290-3139 (e-mails: saad@ee.bilkent.edu.tr, gezici@ee.bilkent.edu.tr).} and Sinan Gezici, {\emph{Senior Member, IEEE}}
% \vspace{-0.6cm}
% }}

%\markboth{Journal of \LaTeX\ Class Files,~Vol.~11, No.~4, December~2012}%
%{Shell \MakeLowercase{\textit{et al.}}: Bare Demo of IEEEtran.cls for Journals}
% The only time the second header will appear is for the odd-numbered pages


\maketitle
\thispagestyle{empty}
\begin{abstract}
\pagestyle{empty}
Millimeter-wave (mmWave) networks, integral to 5G communication, offer a vast spectrum that addresses the issue of spectrum scarcity and enhances peak rate and capacity. However, their dense deployment, necessary to counteract propagation losses, leads to high power consumption. An effective strategy to reduce this energy consumption in mobile networks is the sleep mode optimization (SMO) of base stations (BSs). In this paper, we propose a novel SMO approach for mmWave BSs in a 3D urban environment. This approach, which incorporates a neural network (NN) based contextual multi-armed bandit (C-MAB) with an epsilon decay algorithm, accommodates the dynamic and diverse traffic of user equipment (UE) by clustering the UEs in their respective tracking areas (TAs). Our strategy includes beamforming, which helps reduce energy consumption from the UE side, while SMO minimizes energy use from the BS perspective. We extended our investigation to include \emph{Random}, \emph{Epsilon Greedy}, \emph{Upper Confidence Bound (UCB)}, and \emph{Load Based} sleep mode (SM) strategies.
We compared the performance of our proposed C-MAB based SM algorithm with those of \emph{All On} and other alternative approaches. Simulation results show that our proposed method outperforms all other SM strategies in terms of the $10^{th}$ percentile of user rate and average throughput while demonstrating comparable average throughput to the \emph{All On} approach. Importantly, it outperforms all approaches in terms of energy efficiency (EE).
\pagestyle{empty}

\textit{Index~Terms}--- Beam-forming, contextual MAB, mmWave, reinforcement learning, sleep mode optimization.
\end{abstract}

%\begin{IEEEkeywords}
%Channel switching, jamming, Nash equilibrium, capacity, time-sharing, power allocation.
%\end{IEEEkeywords}

%\IEEEpeerreviewmaketitle

\vspace{-0.3cm}

\section{Introduction}\label{sec:intro}
\pagestyle{empty}
The exponential growth in cellular data demand necessitates an increasing amount of spectrum and has spurred rapid expansion of mobile network infrastructure in recent years. Millimeter wave (mmWave) communications %, operating within the 30-300 GHz band, 
have emerged as a promising technology in fifth-generation (5G) cellular networks. Offering substantial bandwidth, mmWave networks present a viable solution to the pressing issue of spectrum scarcity \cite{shokri2015millimeter}, \cite{baldemair2015ultra}. However, mmWave signals are susceptible to blockage and experience considerable attenuation. To mitigate propagation loss, mmWave BSs are densely deployed with inter-site distances in the order of hundreds
of meters \cite{baldemair2015ultra}, are equipped with large antenna arrays, and utilize efficient spatial multiplexing.

The primary source of power consumption in these BSs is the radio frequency (RF) chain. While the deployment of a large number of RF chains within a BS can be mitigated by combining analog precoding with digital precoding \cite{alkhateeb2015limited} (i.e., hybrid precoding), the energy consumption remains significant. Analog-to-digital converters (ADCs) and digital-to-analog converters (DACs) in an mmWave BS necessitate a considerably higher sampling rate compared to sub-6 GHz systems due to their operation at higher frequencies with larger bandwidths. Given that the power consumption of an ADC/DAC is proportional to the sampling rate, an RF chain in a mmWave BS consequently consumes a substantial amount of energy \cite{feng2018dynamic}.
These factors result in substantial energy consumption, raising both economic and environmental concerns. Specifically, the Information and Communication Technology (ICT) industry is projected to account for approximately 23\% of the global carbon footprint and about 51\% of global electricity consumption by 2030~\cite{andrae2015global}. Addressing this issue through energy-efficient wireless communication has been a significant research focus for over a decade. 

In cellular networks, BSs account for 60\% to 80\% of total energy consumption~\cite{marsan2009optimal}, and interestingly, BS traffic load is less than a tenth of the peak value for 30\% of the time on weekdays \cite{oh2011toward}. This presents an opportunity for energy reduction through dynamic sleep mode optimization (SMO) (i.e., turning on and off BSs). The strategy is to put a BS into sleep mode (SM) when it is serving fewer user equipment (UEs) and shift its load to a nearby BS. However, this can affect the network coverage and decrease the user performance (i.e., throughput, delay). To optimize the balance between energy use and system throughput, we must examine the correlation between these two metrics, and use this information to optimize the BS activation and deactivation. Formulating the optimal BS on/off is, however, a significant challenge. It demands insights such as the number of UEs a specific BS is serving, the resources they are using, how their position will evolve in time, to which cell they will handover, etc. However, such information is typically not available and hard to predict in real-world scenarios, making the task even more difficult.

% \subsection{Contributions}
In this paper, we provide a novel learning-aided SMO that aims to conserve energy while maximizing the overall system throughput. Our framework uses reinforcement learning (RL) and takes into account key aspects such as dynamic UE distribution and state-of-the-art beamforming. Importantly, we trained and tested our RL model over near real-world conditions using advanced modeling. To the best of our knowledge, this is the first study that employs a 3D map for the SMO of mmWave networks. Our solution comprises two phases: The first phase involves strategically deploying BSs within a 3D environment. Initially, numerous potential BSs are randomly placed, followed by selecting a subset of them to maximize spatial coverage across the entire area. This subset will be utilized in the subsequent phase to develop an SMO strategy. The key contributions of this paper include the following:
\begin{itemize}
    \item Incorporating a 3D urban environment model for mmWave communications, capturing real-life scenarios. 
    \item Utilizing beamforming techniques in conjunction with the SMO approach to enhance mmWave link budget, thereby reducing energy consumption for UE.
    \item  The Proposed approach differs from existing literature where multi-armed bandits (MABs) algorithms are used, which suffer from scalability and cannot account for crucial contextual information. UEs are clustered into their corresponding tracking areas (TAs), and this information is utilized as context for the proposed algorithm. Our proposed neural network (NN)-based contextual multi-armed bandit (C-MAB) referred to as \emph{NN-based C-MAB} algorithm effectively manages larger state and action spaces by leveraging NN and contextual information.
    \item Explored the effectiveness of the proposed algorithm by comparing it with other SM strategies: \emph{Random}, \emph{Epsilon Greedy}, \emph{Upper Confidence Bound (UCB)}, \emph{Load Based} \cite{celebi2019load}, and the \emph{All On} \footnote{\emph{All On} approach, can be considered a special case of SM where no BSs are put into SM.}.  
\end{itemize}
The subsequent sections of the paper are structured as follows. Section II provides a comprehensive review of the existing literature. Section III describes the system model and problem formulation. Section IV presents our proposed algorithm for SMO and other SM strategies. Section V details the numerical results and analysis. Finally, conclusions are drawn in Section VI.
% \textcolor{green}{ This approach offers a promising solution for managing mmWave networks in real-world scenarios.}
 % Explored the effectiveness of the proposed algorithm by introducing three novel SM strategies: \emph{Random}, \emph{Epsilon Greedy}, and \emph{UCB}, enriching the evaluation process. Compared the proposed algorithm with an existing \emph{Load Based} \cite{celebi2019load} approach and also included the \emph{All On} approach to see how they all stack up against each other.
%\comment{\cite{guo2013optimal}}
\section{Literature Review}
Despite the extensive research on SMO/on-off scheduling of BSs, a detailed investigation of SMO in mmWave under diverse BS and UE distributions is not available. BS sleep cycles vary in duration: slow cycles last minutes to hours, while fast ones span seconds to minutes \cite{marsan2009optimal}--\nocite{oh2011toward}\cite{celebi2019load}. Queueing theory has been extensively studied to strike a balance between energy savings and the degradation of Quality of Service (QoS) \cite{niu2012energy}, \cite{guo2013optimal}. In \cite{garcia2020energy}, the authors introduced energy-efficient sleep SM techniques for cell-free mmWave massive MIMO networks. They simplified the scenario by modeling the UE locations using a log-normal distributed traffic map. However, this approach, while making the problem more tractable, may not accurately reflect real-world conditions.

Deep learning (DL) and RL have been effectively applied to enhance wireless communication systems. The study in \cite{soorki2017millimeter} investigates a joint stochastic problem related to the placement of BSs and beam steering, aiming to optimize mmWave network coverage. It assumes that the network is aware of the user's position beforehand, while the user's orientation is subject to stochastic changes. However, this assumption may not mirror real-world scenarios accurately, as a user's location can vary considerably over time. Moreover, the study does not directly address the reduction of energy consumption. In~\cite{ho2017energy}, the authors focus on UE power consumption in mmWave M2M communication using beam-aware discontinuous reception (DRX). However, it overlooks BS power usage and assumes periodic beamforming, potentially leading to inefficient resource allocation due to varying data transmission needs.
Study \cite{maghsudi2016multi}, used a single-agent multi-armed bandit (SA-MAB) for small cell activation by a macro BS. However, it did not analyze user distribution.  Similarly, the MAB approach is employed for spectrum scheduling in \cite{li2020multi}. Both methodologies, however, face scalability issues as the complexity of the MAB problem increases with the number of arms, making it difficult to identify the optimal solution. 

SM operation based on deep Q-network (DQN) has been explored in~\cite{liu2018deepnap}, for wireless local-area networks (WLANs). The interrupted Poisson process (IPP) model was employed to simulate bursty traffic patterns. However, the study did not include an analysis of the system’s overall throughput as a result of the SM. The IPP model may not suit all cases, especially mmWave scenarios that depend heavily on line-of-sight (LOS). A significant portion of prior research heavily depends on idealistic assumptions about user movement, such as Poisson traffic models and threshold-based scheduling. However, some of this work encounters scalability issues. Given the complexity of real-world traffic, there is a need for a more flexible and model-free approach capable of solving highly complex problems.

%\includecomment{SMO is a decision-making process where actions are based on past experiences and current states. RL has been widely applied to address such problems, making it particularly suitable for managing BS SMO, especially under real-world traffic conditions that simple models cannot accurately capture.}

%---------------------------------------------------------------------------------------------------------

\vspace{-0.1cm}

\section{System Model}\label{sec:system}
\begin{figure}
\vspace{-0.5cm}
	\includegraphics[width=0.95\linewidth]{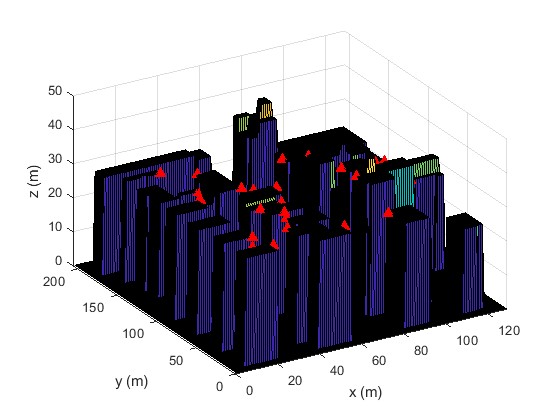}
	\centering
\vspace{-0.3cm}
	\caption{Urban Macro (UMa) outdoor-to-outdoor communication scenario, where the reduced BS positions \eqref{RBS} are represented by red triangles.}
	\label{fig1}
\vspace{-0.35cm}
\end{figure}
Our study is based on an urban macro (UMa) outdoor-to-outdoor communication scenario. This scenario incorporates various elements such as terrain data, landmarks, and street routes, all of which are illustrated in Fig. \ref{fig1}. Note that free 3D geographic data of some real environments can be obtained from publicly available OpenStreetMap (OSM) \cite{haklay2008openstreetmap}, to represent some real-world environments. Using this data, we have constructed a 3D map with a Digital Elevation Model (DEM). This map is structured as a grid, where each square contains data about the location and height of its central point. The model includes buildings of varying heights and widths selected within the ranges of 8-25 meters and 20-45 meters, respectively. The BSs are positioned on the rooftops, while the UEs are located at ground level. We focused on outdoor communication due to the nature of cellular mmWave networks, therefore, we have defined the service area (SA) as the total area excluding the area covered by the buildings. This approach ensures that our focus remains on the areas that require service.

The $N_c$ candidate BS's locations are represented by the set which represents the boundaries of the building facing the SA, 
\begin{equation} \label{PCBS}
\mathcal{P}_{\rm c}^{\mathrm{BS}} \in\left\{\left(x_i^{\mathrm{BS}}, y_i^{\mathrm{BS}}, z_i^{\mathrm{BS}}\right) \mid \forall i \in[N_{\rm c}]\right\}.
\end{equation}

Given the potentially large value of $N_c$, it becomes impractical to position BSs at every building and their respective boundaries. Moreover, simply placing an excessive number of BSs might yield promising results, but this approach does not reflect the real-life scenario where BSs are strategically positioned. In our study, we first strategically minimized the BS candidate locations in a UMa scenario. We used an iterative algorithm that selects the BS with the maximum visibility. Visibility refers to the extent to which a grid is observable from a given location, which is calculated using voxel viewshed algorithm \cite{messerli2015image}. The iterative algorithm adds the BS that improves visibility the most in each iteration. This process continues until no further visibility improvement is possible, resulting in a reduced set of BSs that optimally covers the whole SA, and the reduced collection of BSs positions is denoted by the set
\begin{equation} \label{RBS}
\mathcal{P}_{\rm r}^{\mathrm{BS}} \in\left\{\left(x_i^{\mathrm{BS}}, y_i^{\mathrm{BS}}, z_i^{\mathrm{BS}}\right) \mid \forall i \in[N_{\rm r}]\right\}~.
\end{equation}
where $N_{\rm r}$ (i.e., $N_{\rm r} \ll N_{\rm c}$) represents the total number of BS locations that are capable of covering the entire SA. The SA is split into $M$ grid points where the UEs can be located, and the positions of these points are defined as follows:
\begin{equation} \label{PCUE}
\mathcal{P}^{\mathrm{SA}} \in\left\{\left(x_i^{\mathrm{SA}}, y_i^{\mathrm{SA}}, z_i^{\mathrm{SA}}\right) \mid \forall i \in[M]\right\}~.
\end{equation}

The received signal strength of user $i$ from BS j can be expressed as:
\begin{equation}
P^{\mathrm{rcv}}_{i,j}=P^{\mathrm{tx}}_{j}+ \Upsilon_{i,j}+\PL_{i,j}~.
\end{equation}
where $P^{\mathrm{tx}}_{j}$ denotes the transmit power of $j^{th}$ BS, $\Upsilon_{i,j}$ represents the directivity gain due to beamforming, and $\PL_{i,j}$ is the path loss between user $i$ and BS $j$. The path loss models for the UMa scenario, considering both LOS and non-line of sight (NLOS) conditions, have been utilized as outlined in the 3GPP TR 38.901 technical report \cite{zhu20213gpp}:
\begin{equation}
\PL_{i, j} = \left\{\begin{array}{l}
28.0 + 20\log_{10}(f_{\rm o}) + 22\log_{10}(d^{3D}_{i,j}), \text { for LOS } \\
32.4 + 20\log_{10}(f_{\rm o}) + 30\log_{10}(d^{3D}_{i,j}), \text { for NLOS }
\end{array}\right.\nonumber 
\end{equation}
where $d^{3D}_{i,j}$ is the 3D distance (in meters) between the BS $j$ and the UE $i$, and $f_{\rm o}$ is the center frequency normalized by 1~GHz.

We defined binary variables $u_{i,j}$, and $s_{i,j}$ to represent the user association to BS, and the presence of a user within the SA of other BSs except the serving BS, respectively:
\begin{equation}
\begin{aligned}
u_{i, j} & \doteq\left\{\begin{array}{l}
1, \text { user } i \text { is served by } \mathrm{BS}  j \\
0, \text { otherwise }
\end{array},\right. \\
s_{i, j} & \doteq\left\{\begin{array}{l}
1, \text { user } i \text { is in SA of } \mathrm{BS}  j \text { and } u_{i, j} \neq 1\\
0, \text { otherwise }
\end{array},\right. \\
\end{aligned}\nonumber 
\end{equation}
where $i =1,2, \ldots, U, j=0,1, \ldots, N$, (i.e., $U\leq M$, $N\leq N_{\rm r}$). $U$ is the total number of users that can be located anywhere within the SA, as defined by the set $\mathcal{P}^{\mathrm{SA}}$. On the other hand, $N$ denotes the number of BSs whose locations are selected from the reduced set $\mathcal{P}_{\rm r}^{\mathrm{BS}}$ ($\mathcal{P}_{\rm r}^{\mathrm{BS}} \subseteq \mathcal{P}_{\rm c}^{\mathrm{BS}}$). The binary variable $s_{i, j}$ quantifies interference from non-serving, nearby BSs. 

The throughput of the user $i$ can be calculated as:
\begin{equation}
\begin{aligned}
T^{\mathrm{put}}_i=B_i^w\log_2\left(1+\frac{\sum_{k=1}^{N} u_{i, k} P^{\mathrm{rcv}}_{i,k} }{\sum_{k=1}^{N} s_{i, k} P^{\mathrm{rcv}}_{i,k} +\psi_i}\right)
\end{aligned}~.
\end{equation}
where $B_i^w$ denotes the bandwidth allocated to user  $i$, and the noise $\psi_i$ is given by $\psi_i=\Gamma\nu B_i^w\xi$. In this equation, $\Gamma$ is Boltzmann’s constant, $\nu$ represents the temperature, and, $\xi$ is the noise figure which quantifies the degradation of the signal-to-noise ratio (SNR). The product of $\Gamma$,  $\nu$, $B_i^w$ represents the thermal noise and $\xi$ represents the impact of the noise figure on the throughput assigned to user $i$.

Energy efficiency (EE) is a crucial performance metric that quantifies the amount of energy consumed per received information bit. It can be mathematically represented as:
\begin{align}
  \EE=\frac{ \sum_{i=1}^{U}T^{\mathrm{put}}_i}{\sum_{j=1}^{N} P_{\gNB}^{j}}~.
\end{align}

In a mmWave network, the power consumed by the $j^{th}$ BS, denoted as $ P_{\gNB}^{j}$, is a combination of the power consumed by the Base Band Unit (BBU) and the Active Antenna Unit (AAU). This is described in \cite{gao2021energy} as follows: 
\begin{equation}
       P_{\gNB}^j = \frac{{(P_{{{\text{BBU}}}} + P_{{{\text{AAU}}}} )}}{{(1 - \varrho_{{{\text{cooling}}}} )(1 - \varrho_{{{\text{DC}}}} )}}.
\end{equation}
Here, $\varrho_{{{\text{cooling}}}}$ and $\varrho_{{{\text{DC}}}}$ represent the power consumption of the cooling module and DC conversion loss, respectively. The power consumed by the BBU and AAU are denoted by $P_{{{\text{BBU}}}}$ and $P_{{{\text{AAU}}}}$, respectively. The values for the power consumption of different modules are taken from \cite{gao2021energy}.

% \vspace{-0.22cm}
\subsection{Problem Formulation}
The primary goal of this research is to transition the BS into SM in a manner that minimally impacts the system’s overall throughput. By doing so, we can conserve energy from both the BS and UE perspectives. This is because putting the BS into SM helps to narrow down the beam search space, thereby reducing the power consumption of the UE. The operational status of the BS is represented by a binary variable $\gNB_{j} \in \{0,1\}$, where $j=1,2,\dots,N$. A value of 1 indicates that the BS is on, while 0 signifies that it is off. The objective function can be formulated as:
% \underset{\{\nu,P_1,P_2,\Po_1,\Po_2\}}
\vspace{-0.1cm}\begin{subequations}\label{eq:obj}
\begin{align}\label{eq:obj_a}
\max~& \sum_{i=1}^{U}T^{\mathrm{put}}_i \\\label{eq:obj_b}
{\rm{subject~to}}~ &\sum_{i=1}^{U} u_{i,j}=1\,,\\\label{eq:obj_c}
&N-\sum_{j=1}^{N} \gNB_{j}=  \lfloor \alpha_{\mathrm{off}} N \rfloor
\,~.
% &P_1\in[0,\Ppk],\,P_2\in[0,\Ppk]\,,
% \\\label{eq:probhmax_e}
% &\Po_1\in[0,\textcolor{blue}{\Pop}],\,\Po_2\in[0,\textcolor{blue}{\Pop}],\,
% %\\\label{eq:probhmax_f}
% \nu\in[0,1]\,,
\end{align}
\end{subequations}
The optimization problem in \eqref{eq:obj} is complex due to unknown user distribution and varying BS load. Constraint \eqref{eq:obj_b} ensures only one BS serves a UE, while \eqref{eq:obj_c} forces $\lfloor \alpha_{\mathrm{off}} N \rfloor $ BSs to be deactivated. %considering the blockages, UEs traffic, and propagation issues in mmWave BSs, 
$\alpha_{\mathrm{off}}$ is the SM activation parameter, which represents the percentage of BS to be turned off, where $0 \leq \alpha_{\mathrm{off}} \leq1 $. 

% \begin{equation}
% \begin{aligned}
% SBS_{j} & \doteq\left\{\begin{array}{l}
% 1, \mathrm{BS_j} \text { is ON }  \\
% 0, \text { otherwise },
% \end{array}\right. \\
% j=0,1, \ldots, N.
% \end{aligned}
% \end{equation}
% \vspace{-0.05cm}
%\section{Contextual Multi-Armed Bandits Formulation for SMO}
\section{Sleep Mode Strategies}
In this section, we elaborate on our proposed \emph{NN-based C-MAB} SMO algorithm. Additionally, we defined alternative SM strategies such as \emph{Load Based}, \emph{Random}, \emph{Epsilon Greedy}, \emph{UCB}, and \emph{All On}, with which we will compare the performance using computer simulations.

\subsection{NN-Based C-MAB Formulation for SMO}
The optimization problem \eqref{eq:obj} discussed earlier is non-convex and falls into the category of NP-hard problems, which are particularly challenging to solve. Traditional methods, while useful, often make idealized assumptions that can lead to a loss of precision. However, the introduction of deep RL has provided a more robust approach, capable of handling more realistic assumptions and managing the nonlinear mapping from the state/context space to the action space. To address the complexities and inherent difficulties of this non-convex problem, we propose the use of a \emph{NN-based C-MAB} approach. 

\begin{figure}
\vspace{-0.5cm}
	\includegraphics[width=0.98\linewidth]{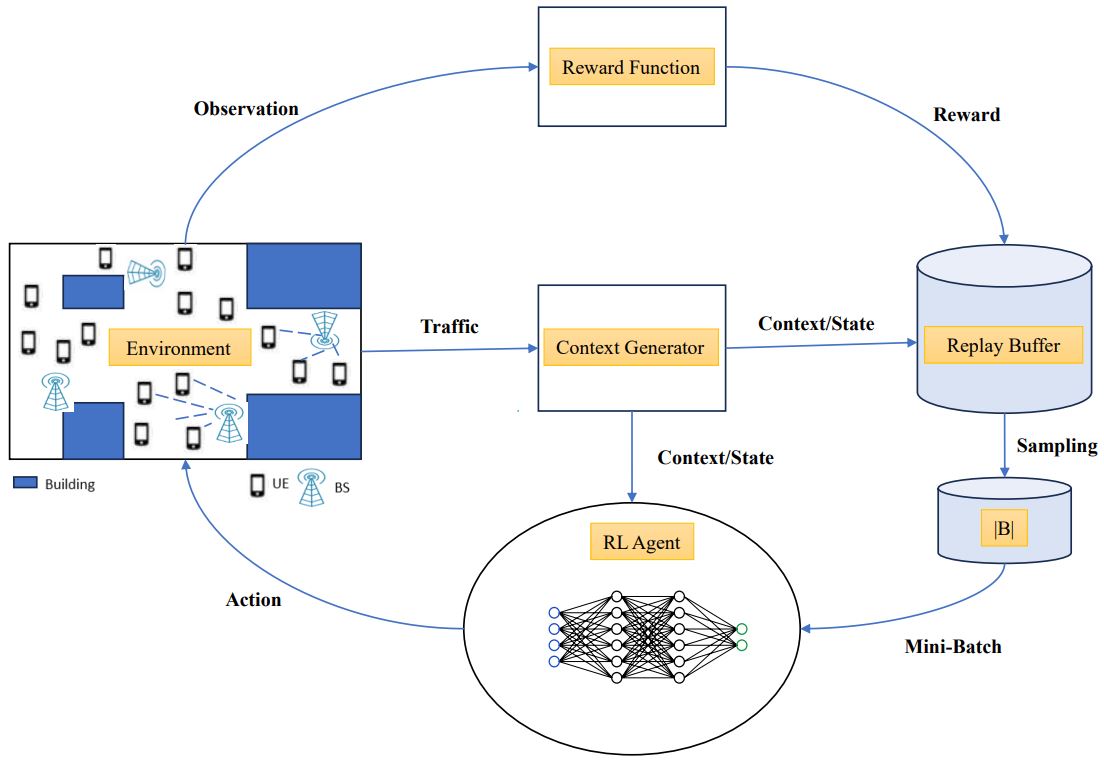}
	\centering
\vspace{-0.3cm}
	\caption{The proposed RL framework for SMO in mmWave networks.}
	\label{fig2}
\vspace{-0.5cm}
\end{figure}

To formulate the SMO for mmWave as an RL task, we represent the system as a combination of five components, as shown in Fig. \ref{fig2}: 1) an environment emulator, which encapsulates the behavior of the BS and UEs; 2) a context generator, which helps the agent in taking action based on the environment; 3) an RL agent, which reacts to the context and takes an action; 4) a reward function, which emits a scalar reward based on the action taken by the agent; and 5) a replay buffer, which serves as the data plane to update the belief of the RL agent. In each iteration t, the context generator block produces a context $\mathbf{c}_t$, for the RL agent, which takes an action $\mathbf{a}_t^{\text{C-MAB}}$. The reward function emits a reward $r_t$, based on the action taken, which is then stored in the replay buffer with the context. A batch $\mathbf{B}$, from the buffer is used to update the RL agent's weights. This process optimizes the SM for mmWave systems, ensuring network efficiency.

C-MAB \cite{lu2010contextual}, enables the agent to choose future actions based on contextual information, actions (i.e., arms), and rewards learned from past observations. An NN is incorporated into the C-MAB framework to effectively manage the infinite state space and enhance decision-making based on diverse contexts, while the exploration dilemma is handled using an epsilon decay over time, and NN is responsible for exploiting the best action. Unlike the MAB algorithm, which selects actions independently of the environment's state, the C-MAB algorithm tailors its decisions to the observed context, thereby enabling a more personalized approach to each situation. Upon observing the context, the C-MAB algorithm selects an action that yields a reward. The overarching objective of this algorithm is to maximize the cumulative reward, thus optimizing the decision-making process.

In the context of the MAB algorithm, each round allows the agent to perceive the reward of the chosen arm. The agent's objective is to reduce the regret, defined as the discrepancy between the chosen arm $\rho_{\rm c}(t)$ and the optimal arm $\rho^*(t)$ that could have been selected. The cumulative regret $\regret$ over the entire time horizon is given by:
\begin{equation}
\regret=\mathbb{E}\left[\sum_{t=1}^T \rho^*(t)-\sum_{t=1}^T \rho_{\rm c}(t)\right] ~.
\end{equation}
\noindent where expectation accounts for the algorithm's randomness and the disclosed rewards.
In the MAB problem, the goal of the proposed algorithms is to balance exploration and exploitation. Exploration assesses potential arms, while exploitation maximizes reward from the best arm.
%\comment{An algorithm overly focused on exploration may not utilize the best arm frequently, leading to sub-optimal rewards. Conversely, an algorithm that primarily exploits without sufficient exploration may become trapped at a sub-optimal arm. There are numerous algorithms, such as Epsilon-Greedy and UCB, that strive to balance exploration and exploitation for optimal long-term rewards.}
% \vspace{-0.5cm}
\subsubsection{NN-based C-MAB}
In the SMO framework, we employ an \emph{NN-based C-MAB} combined with an epsilon decay strategy to balance exploration and exploitation. In the early stages of learning, the model faces a high degree of uncertainty due to limited data, necessitating broad exploration of the environment. As the model accrues more data, it gains a deeper understanding of the environment, reducing uncertainty and shifting the focus towards exploiting the optimal action to maximize cumulative reward. Consider a tuple ($\mathcal{C}$, $\mathcal{A}$, R) representing the core elements of an RL process. Here, $\mathcal{C}$ is the context or state space with environmental data, $\mtA$
 is the action space with $A_{\text{total}}$  possible actions, and R is the reward function. An RL agent interacts with an environment to model $P(r|a,c)$, observing a context $\mathbf{c}_t \in \mathcal{C}$ at each time step $t$. The agent selects an action $\mathbf{a}_t \in \mtA$ and receives a reward $r_t=R(\mathbf{a}_t^{\text{C-MAB}},\mathbf{c}_t)$ based on the chosen action in the observed context. However, the agent cannot observe the reward from unchosen actions. The agent's objective is to maximize the cumulative reward $\Lambda_t=\sum_{t=1}^{T}\kappa^{t-1}r(\mathbf{a}_t^{\text{C-MAB}},\mathbf{c}_t)$, where $\kappa$ is the discount factor. 
The design of these essential elements significantly impacts the algorithm’s convergence. 
\begin{itemize}
    \item \textbf{Action space} $\mtA$: To formulate the SMO using \emph{NN-based C-MAB}, we defined the action space as follows:
\begin{equation}
\begin{aligned} \label{Action Space}
\mathbf{\mtA} &= \Big\{ \gNB_{1} \in \{0,1\}, \ldots, \gNB_{N} \in \{0,1\} \mid \\
&\quad \sum_{j=1}^N \gNB_{j} = \left\lfloor \alpha_{\mathrm{off}} N \right\rfloor\Big\}~.
\end{aligned}
\end{equation}

In time slot $t$, action $\mathbf{a}_t^{\text{C-MAB}}$ $\in$ $\mtA$ switches the BS station between sleep and active modes, and the action vector $\mathbf{a}_t^{\text{C-MAB}}$ is defined as:
\begin{align}
\resizebox{0.9\columnwidth}{!}{$
\mathbf{a}_t^{\text{C-MAB}}=\left[\gNB_1\in\{0,1\}, \gNB_2\in\{0,1\},\cdots, \gNB_N\in\{0,1\}\right]$}~.\nonumber
\end{align}
The total number of actions is represented by:
\begin{equation}
\begin{aligned}
A_{\text{total}} = \binom{N}{\left\lfloor \alpha_{\mathrm{off}} N \right\rfloor}.
\end{aligned}
\end{equation}

\item \textbf{Context/State space} $\mathcal{C}$: The context at each time step $t$ is generated using $K$-means clustering, which groups UE based on their TA. The exact location of the UE is unknown to the BS due to privacy concerns, and it is computationally expensive to determine the exact location of the BS to which the UE is connected. However, the system is aware of the TA in which the UE is located, and it updates whenever the UE moves to a new TA. The UE is clustered into $K$ clusters based on their TA. The coordinates of the clusters and the density of each cluster from the context can be defined as follows:
\begin{align}
\resizebox{0.9\columnwidth}{!}{$
\mathbf{c}_t=[(dx_1,dy_1), (dx_2,dy_2),\cdots, (dx_K,dy_K), \mu_1, \mu_2,\cdots, \mu_K]
$},\nonumber
\end{align}
where, $dx_i$, and $dy_i$ represent the center $x$ and $y$ coordinates of the clusters $i \in(1, \cdots, K)$, respectively, and $\mu_i$ represents the proportion of UEs in cluster $i$ to the total number of UEs in the entire area.
\item \textbf{Reward}: We choose the reward at a given time step $t$, represented as $r_t(\mathbf{a}_t^{\text{C-MAB}},\mathbf{c}_t)$, to be the $10^{th}$ percentile of the total user throughput which promotes fairness by improving experiences for users with the lowest throughput:
\begin{align}
\label{RewardFunciton}
r_t(\mathbf{a_t^{\text{C-MAB}*}},\mathbf{c}_t) = \mathrm{{Percentile}}_{10}\left(\sum_{i=1}^{U}T^{\mathrm{put}}_i\right) ~.
\end{align}
where $\mathbf{a_t^{\text{C-MAB}*}}$ is the action taken at time $t$ out of $A_{\text{total}}$ (i.e., $A_{\text{total}}= |\mtA|$) possible actions. This reward function is robust against outliers and motivates overall network performance enhancement. 
\end{itemize}

% Consider a tuple (C, A, R) representing the core elements of a Markov Decision Process (MDP). Here, C is the context or state space with environmental data, A is the action space, and R is the reward function. The reward $\mathcal{R}(c,a)$ corresponds to a specific environment and agent action. The agent's policy at time slot $t$ aims to maximize the cumulative discounted reward, denoted as $\Lambda_t=\sum_{t=1}^{T}\kappa^{t-1}r(\mathbf{c_t},\mathbf{a_t})$.

A neural network with $L$ hidden layers is employed to model $P(r_t|\mathbf{a_t}^{\text{C-MAB}},\mathbf{c_t}, \Theta)$, where $\Theta$ denotes the model's weights. The network is trained via an Adam optimizer, incorporating L2 regularization ($\lambda$). At each time step $t$, the context is input to the network $\mathbf{c}_t$, where $t \in {1,2,\cdots,T}$. The action yielding the highest expected reward is selected. Updating the model after every iteration can be computationally expensive, resulting in noisy updates, overfitting, and an inability to handle concept drifts. Therefore, the model is updated every $\tau_{\text{{update}}}$ iterations using a randomly chosen batch of size $|\textbf{B}|$ from the replay buffer.

At each iteration, a random action is selected with a probability of $\epsilon^{\text{C-MAB}}$, while the action chosen by the NN is selected with a probability of $(1-\epsilon^{\text{C-MAB}})$. The value of $\epsilon^{\text{C-MAB}}$ decreases with each iteration according to the formula $\epsilon^{\text{C-MAB}} = \epsilon^{\text{C-MAB}} * \epsilon_{\mathrm{th}}^{\text{C-MAB}}$, where $0 < \epsilon_{\mathrm{th}}^{\text{C-MAB}}<1$ to control the rate of decay.  A high value (close to $1$) means slow decay and more exploration, while a low-value %(close to 0)
means fast decay and quicker transition to exploitation. This strategy ensures that the model explores adequately in the early stages and gradually transitions towards exploitation as its knowledge base expands.

\subsection{Load Based SM Strategy \cite{celebi2019load}}
In the Load-based algorithm, the BS with the lowest load is designated to enter SM. The load factor for each UE \(i\) is defined as follows:
\begin{equation}
L_{i}^\text{UE} = \left\{\begin{array}{l}
\frac{1}{\sum_{j=1}^N u_{i, j}+s_{i, j}},  \text{if } u_{i, j} \text{ or } s_{i, j} > 0
 \\
0, \text { otherwise }
\end{array}\right.~,\nonumber 
\end{equation}
where $\frac{1}{\sum_{j=1}^N u_{i, j}+s_{i, j}}$ denotes the count of BSs from which a UE can receive service. Based on this, the load value of  \(j^{th}\) BS can be defined as the sum of load factors of UEs associated with the \(j^{th}\) BS, as follows: 
\begin{equation}
L_j^{\text{BS}}=\sum_{i=1}^U u_{i,j}L_{i}^\text{UE}~.
\end{equation}
Under this algorithm, a total of \(\lfloor \alpha_{\mathrm{off}} N \rfloor\) BS will be turned off. Notably, these are the BS characterized by the minimum load values $L_j^{\text{BS}}$.

\subsection{Random SM Strategy}
In the \emph{Random} SM Strategy, a total of \(\lfloor \alpha_{\mathrm{off}} N \rfloor\) BS are independently selected for SM. Employing this random approach facilitates a fair comparison with other strategies, as it serves as a baseline for evaluating the effectiveness of more sophisticated algorithms.

\subsection{UCB SM strategy}
The \emph{UCB} bandit algorithm \cite{masrur2023outage}, a state-of-the-art MAB algorithm, addresses the challenges posed by non-stationary environments. It emphasizes the importance of exploring various actions while also exploiting the most promising ones to maximize total rewards. This strategy embodies the principle of `optimism in the face of uncertainty', striking a balance between exploration and exploitation for optimal performance.
The action space is similar to that of our \emph{NN-based C-MAB}, defined in \eqref{Action Space}.  
% \begin{equation}
% \begin{aligned}
% \mathbf{\mtA} &= \{ \gNB_{1} \in \{0,1\}, \ldots, \gNB_{N} \in \{0,1\} \mid \\
% &\quad \sum_{j=1}^N \gNB_{j} = \left\lfloor \alpha_{\mathrm{off}} N \right\rfloor\}
% \end{aligned}
% \end{equation}
Mathematically, the \emph{UCB} algorithm selects an action $\mathbf{a_t}$ at iteration $t$ with the highest upper confidence bound, given by:
\begin{equation}
    \mathbf{a_t^{UCB}} =\underset{k\in\{1,\ldots,A_{total}\}}{\arg \max }\left( \omega_k + \delta \sqrt{\frac{2 \ln(t)}{n_k}} \right)~.
\end{equation}
where, \(\omega_k\) and \(n_k\) represent the average reward obtained from action \(k \in \mathcal{A}\) and the number of times action \(k\) has been selected, respectively.
\subsection{Epsilon Greedy SM Strategy}
The \emph{Epsilon Greedy} SM approach balances exploration and exploitation using a parameter, epsilon \(\epsilon^{\text{greedy}}\), to regulate the choice between random selection and maximizing expected rewards. It randomly selects actions with probability \(\epsilon^{\text{greedy}}\) and prioritizes actions with higher average rewards with probability 1-\(\epsilon^{\text{greedy}}\).  
\subsection{All On Strategy}
In the \emph{All On} Strategy, as the name suggests, no BSs are turned into SM. This strategy serves as a crucial benchmark for evaluating the effectiveness of other SM strategies. It provides a baseline comparison to assess the impact of SM strategies on network performance.

\section{Numerical Results and Analysis}\label{sec:Nume}

In this section, we evaluate the efficiency of the proposed SMO using a 3D model that simulates a UMa outdoor-to-outdoor communication scenario, as depicted in Fig. \ref{fig1}. The 3D model, focusing on outdoor users, spans an area of $129~{\rm m}\times 206~{\rm m}\times  45~{\rm m}$ ($x, y, z$, respectively), with the UE height set at $1.5$~m above the ground level. The entire area is divided into a grid with a resolution of $1~{\rm m}\times 1~{\rm m}$. The BSs are positioned at the edges of buildings, excluding those near the boundary of the area, resulting in a total of $N_{\rm c}=143$ BS candidate locations, denoted as $\mathcal{P}_{\rm c}^{\mathrm{BS}}$ \eqref{PCBS}. As explained in Section~\ref{sec:system}, we reduce the candidate BS locations to $N_{\rm r}=31$, with locations given by $\mathcal{P}_{\rm r}^{\mathrm{BS}}$ \eqref{RBS}. These BS can provide coverage to the entire SA $\mathcal{P}^{\mathrm{SA}}$ \eqref{PCUE}, which consists of $M=11504$ grid points where a UE could potentially be located. In this study, we consider a system operating at a carrier frequency of $f_{\rm o}=28$~GHz and a transmit power of $P_{\mathrm{tx}}=20$~dBm. The system has a total bandwidth of $50$~MHz and utilizes a round-robin allocation scheme. We use Boltzmann's constant $\Gamma$, valued at $1.38\times 10^{-23}$, and assume a temperature $\nu$ of $298$~Kelvin. The noise figure $\xi$ in our system is $9$~dB.

An NN architecture is utilized, consisting of two hidden layers ($L=2$), with 128 and 64 neurons respectively. The Rectified Linear Unit (ReLU) activation function is applied in the hidden layers, while the output layer uses a linear activation function. L2 regularization is incorporated to prevent overfitting, with a parameter of $\lambda=1 \times 10^{-4}$. For the epsilon decay algorithm, $\epsilon^{\text{C-MAB}}=0.7$ is used for the exploration-exploitation trade-off, and the decay rate $\epsilon_{\mathrm{th}}^{\text{C-MAB}}$ is set to $0.9$. For the \emph{Epsilon Greedy} algorithm, we set \(\epsilon^{\text{greedy}}\) to 0.4, and for the \emph{UCB} algorithm, \(\delta\) is set to 4. To ensure a fair comparison, the reward used by both the \emph{Epsilon Greedy} and \emph{UCB} algorithms is similar to that of \emph{NN-based C-MAB} \eqref{RewardFunciton}. The model weights $\Theta$ are updated after every $\tau_{\text{{update}}}=8$ iterations, and the batch size is set to $|\mathbf{B}|=256$.

In Fig.~\ref{RateViter}, Fig.~\ref{PViter}, and Fig.~\ref{EE}, we conduct experiments where we randomly select \(N=15\) BSs from \(\mathcal{P}_{\rm c}^{\mathrm{BS}}\) and place \(U=70\) UEs randomly in \(\mathcal{P}^{\mathrm{SA}}\) at each iteration \(t\). We assume \(K=10\) TA (equivalent to 10 clusters), K-means clustering is used to cluster UE is their TA. SM activation parameter of \( \alpha_{\mathrm{off}}=0.3\), indicating that 30\% of the BSs will be put into SM.
\subsection{Performance of NN-Based C-MAB}

\begin{figure}
%\vspace{-0.5cm}
	\includegraphics[width=0.9\linewidth]{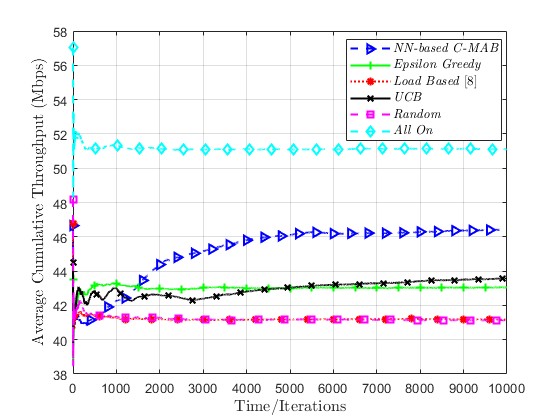}
	\centering
\vspace{-0.3cm}
	\caption{Performance of \emph{NN-based C-MAB} for SMO vs. other methods in terms of throughput considering $N=15$, $U=70$, and $\alpha_{\text{off}}=0.3$.}
	\label{RateViter}
\vspace{-0.35cm}
\end{figure}

%\textbf{Performance of NN-Based C-MAB.}
Fig.~\ref{RateViter} presents a comparison of our proposed \emph{NN-based C-MAB} algorithm with SM  strategies in terms of average cumulative throughput. Our proposed algorithm outperforms all other SM strategies, achieving an average (over the number of users) cumulative throughput of $46.3874$~Mbps, compared to roughly $43$~Mbps for the other approaches. The proposed method rate is close to the \emph{All On} approach (where all BSs are On), which achieves a data rate of $51.1102$~Mbps. % Furthermore, our proposed \emph{NN-based C-MAB} exhibits superior training speed.

\begin{figure}
%\vspace{-0.5cm}
	\includegraphics[width=0.9\linewidth]{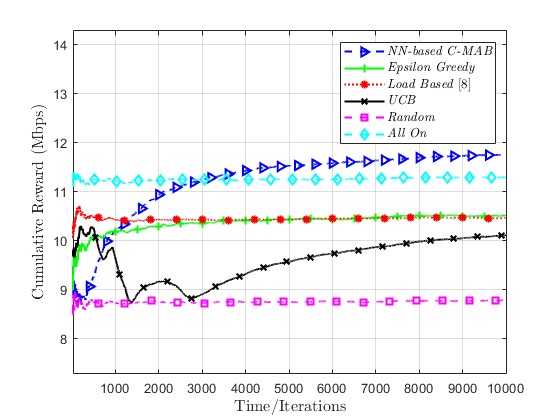}
	\centering
\vspace{-0.3cm}
	\caption{Comparison of $10^{th}$ percentile user rate via \emph{NN-based C-MAB} and other SM strategies with $N=15$, $U=70$, and $\alpha_{\text{off}}=0.3$.}
	\label{PViter}
\vspace{-0.35cm}
\end{figure}

The Fig.~\ref{PViter} illustrates the cumulative $10^{\rm th}$ percentile (cumulative reward) user rate for all methods. Our proposed \emph{NN-based C-MAB} outperforms all other approaches, even the \emph{All On} approach. This superior performance is attributed to effective SMO strategy and interference management. In a dense network, high interference from nearby BSs can degrade signal quality, particularly for UEs at the cell edge or in high interference zones, resulting in lower overall throughput as shown in Fig. \ref{RateViter}. However, turning off a BS alters the interference landscape, leading to reduced interference. Consequently, the rate for the worst 10\% of UEs improves, yielding a higher $10^{\rm th}$ percentile rate. This illustrates a trade-off between total throughput and quality of service, as enhancing the rate for the worst-performing UEs can boost the overall service quality.
%This results in a higher $10^{\rm th}$ percentile rate, demonstrating a trade-off between overall throughput and quality of service.

The random SM strategy performed the worst, as anticipated, across both metrics. Meanwhile, the \emph{Epsilon Greedy}, \emph{Load Based}, and \emph{UCB} approaches showed similar performance. Although the Greedy approach can often perform well, its performance isn't always guaranteed. On the other hand, the \emph{UCB} algorithm in this case might select suboptimal actions due to its ineffective ability to balance between exploration and exploitation as compared to \emph{NN-based C-MAB}. The \emph{Load Based} method is not working well in terms of both throughput and $10^{\rm th}$ percentile. The instantaneous load of a BS in real networks can fluctuate due to blocked calls, association policies, traffic patterns, and transmission rates, making it an imperfect representation of the exact load distribution.
\subsection{Normalized EE Analysis}
\begin{figure}
%\vspace{-0.5cm}
	\includegraphics[width=0.9\linewidth]{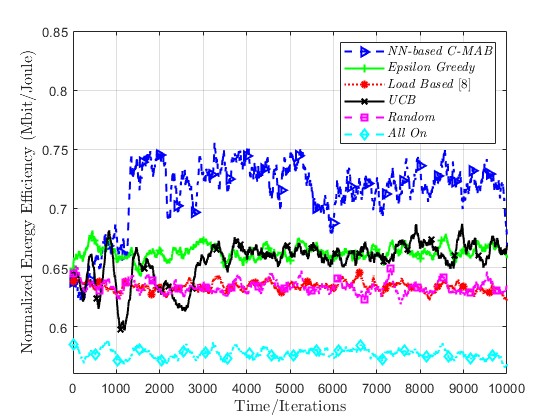}
	\centering
\vspace{-0.3cm}
	\caption{Comparing NEE of the \emph{NN-based CMAB} with other SM strategies, considering $N=15$, $U=70$, and $\alpha_{\text{off}}=0.3$.}
	\label{EE}
\vspace{-0.35cm}
\end{figure}

In addition to the overall system throughput and the $10^{th}$ percentile of user rates, we also evaluate the EE of the system. Fig.~\ref{EE} shows the normalized energy efficiency (NEE) with a moving average calculated over 200 iterations to temper the effect of short-term fluctuations. The proposed \emph{NN-based C-MAB} approach demonstrates superior NEE compared to the other methods. EE is also a function of throughput. The other SM approaches achieved lower throughput compared to the proposed approach, resulting in reduced EE. This indicates that the \emph{NN-based C-MAB} approach excels in resource utilization, achieving higher throughput with less energy.
\subsection{Rate versus Number of Users}
\begin{figure}
%\vspace{-0.5cm}
	\includegraphics[width=0.9\linewidth]{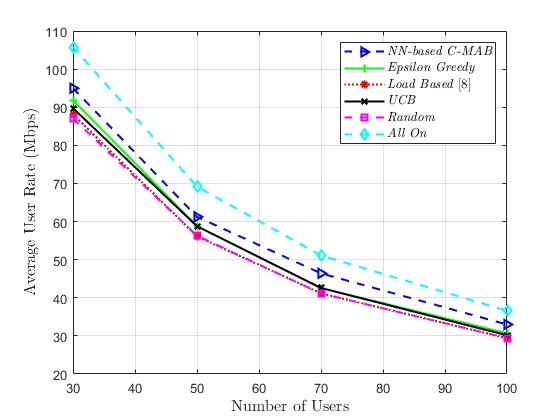}
	\centering
\vspace{-0.3cm}
	\caption{Average user throughput versus the number of UEs ($U$) for a fixed number of BSs ($N=15$) and $\alpha_{\text{off}}=0.3$, comparing the proposed SM strategy with others.}
	\label{rateVsUE}
\vspace{-0.35cm}
\end{figure}

%\textbf{Rate versus Number of Users.} 
Next, we examine a scenario with a fixed number of BSs, denoted as $N$=15, while varying the number of UEs, denoted as $U$, from $30$ to $100$. The average user throughput for this configuration is presented in Fig.~\ref{rateVsUE}, where it is evident that our proposed approach consistently outperforms the other SM strategies, namely \emph{Load Based}, \emph{UCB}, \emph{Epsilon Greedy}, and \emph{Random}. However, as the number of UEs increases, there is a noticeable decrease in the average user throughput. This is attributed to the fact that the number of BSs remains constant, resulting in the same resources being shared among an increasing number of UEs. Consequently, the average throughput decreases.

\subsection{Effect of Increased Action Space on Performance}
\begin{figure}
%\vspace{-0.5cm}
	\includegraphics[width=0.9\linewidth]{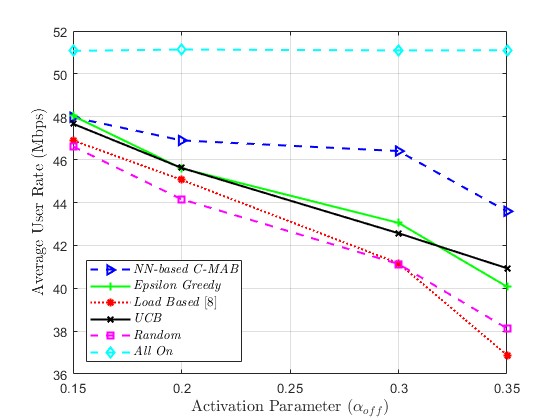}
	\centering
%\vspace{-0.3cm}
	\caption{The average user rates of different SM strategies as the action space increases with varying $\alpha_{\text{off}}$ values, with U=70, and N=15.}
	\label{aciton space}
\vspace{-0.35cm}
\end{figure}

We fixed the number of UEs ($U=70$), and BSs ($N=15$), and varied the $\alpha_{\text{off}}$  value. With $\alpha_{\text{off}}$ rising from 0.15 to 0.35, signifying more BSs being put to sleep, the action space expands from 105 to 3003 actions. As the action space increases with the provided contextual information, the \emph{NN-based C-MAB} approach consistently outperforms other SM strategies (Fig.~\ref{aciton space}). Despite the declining average user rate across all SM strategies except \emph{All On}, attributed to the reduced resource at each step as more BSs are put to SM, the \emph{NN-based C-MAB} maintains its superiority. Notably, it demonstrates adaptability by leveraging contextual information, a feature absent in other approaches. \emph{All On} approach keeps all BSs active, resulting in consistent performance. \emph{UCB} surpasses \emph{Epsilon Greedy} in larger action spaces, showcasing its adeptness in handling the exploration-exploitation trade-off. In contrast, \emph{Epsilon Greedy} performance diminishes due to its simplistic focus on exploitation.
Conversely, \emph{Random} and \emph{Load Based} strategies show inferior performance with increasing action space. 

\section{Conclusion}
In this paper, we study the SMO of mmWave BSs in a UMa 3D propagation environment. Our goal was to achieve high system-wide throughput while reducing the overall network energy consumption. The UEs were randomly placed in the SA at each iteration, simulating the dynamic distribution of users in real-world environments. We also considered interference from other BSs that serve the UEs using beamforming. Due to the dynamic distribution of users leading to an infinite state space, traditional approaches to solving the SMO problem proved to be challenging. Therefore, we addressed the optimization problem using an \emph{NN-based CMAB} algorithm, an RL framework. The UEs were clustered into their respective TAs, which served as the context. NN was then incorporated to map this context to the action, with the additional aid of the epsilon greedy decay algorithm, facilitating the exploration of the environment. To assess the efficiency of the proposed approach, we extensively compared it with various other SM strategies. Numerical results demonstrated the effectiveness of our proposed approach in terms of normalized EE, average throughput, and the $10^{th}$ percentile of the user rate. Furthermore, the numerical results underscore the effectiveness of the \emph{NN-based C-MAB} approach in handling larger action spaces. This work contributes to ongoing efforts to enhance the sustainability of 5G networks, offering a promising solution for managing real-world mmWave networks. Our future work will analyze the impact of other factors on the performance of the proposed C-MAB SMO approach, such as the availability of reflectors and blockages within the environment.
\vspace{-0.2cm}

\bibliographystyle{IEEEtran}
\bibliography{mybib}

\end{document}